\documentclass[a4paper,12pt,onecolumn]{article}
\usepackage[utf8]{inputenc}
\usepackage{geometry}
\usepackage[authoryear,sort]{natbib}
\setcitestyle{open={(},close={)}}
\usepackage{setspace}
\usepackage{multirow}
\usepackage{makecell}
\renewcommand{\baselinestretch}{1.5}
\usepackage{cite}
\usepackage{amsmath,amsfonts,amssymb,amsthm,amsbsy}
\usepackage{enumerate}
\usepackage{booktabs}
\usepackage{color}
\usepackage{graphicx}
\usepackage{subfigure}
\usepackage[ruled,vlined]{algorithm2e}
\usepackage{bbm}
\usepackage{diagbox}
\usepackage{makecell}

\newcommand{\mb}[1]{\pmb{#1}}

\newcommand{\noinhead}[1]{\noindent\textbf{#1}}

\newcommand\refeq[1]{(\ref{#1})}
\newcommand\refalg[1]{Algorithm~{\ref{#1}}}
\newcommand\reffig[1]{Figure~{\ref{#1}}}
\newcommand\reftab[1]{Table~{\ref{#1}}}
\newcommand\refsec[1]{Section~{\ref{#1}}}

\usepackage[hidelinks]{hyperref}

\title{Evaluation and Control of Opinion Polarization and Disagreement: A Review}
\author{Yuejiang~Li and H.~Vicky~Zhao\thanks{Y. Li and H. V. Zhao are with department of automation, Tsinghua University. lyj18@mails.tsinghua.edu.cn, vzhao@tsinghua.edu.cn.}}
\date{}

\begin{document}

\maketitle

\begin{abstract}
\noinhead{Purpose:} With the recent advances of networking technology, connections among people are unprecedentedly enhanced.
People with different ideologies and backgrounds interact with each other, and there may exist severe opinion polarization and disagreement in the social network.
There have been a lot of reviews on modeling opinion formation.
However, less attention has been paid to opinion polarization and disagreement.

\noinhead{Design/methodology/approach:} In this work, we review recent advances in opinion polarization and disagreement and pay attention to how they are evaluated and controlled.

\noinhead{Findings:} In literature, three metrics: polarization, disagreement, and polarization-disagreement index (PDI) are usually adopted, and there is a tradeoff between polarization and disagreement.
Different strategies have been proposed in literature which can significantly control opinion polarization and disagreement based on these metrics.

\noinhead{Originality/value:} This review is of crucial importance to summarize works on opinion polarization and disagreement, and to the better understanding and control of them.
\end{abstract}

\noinhead{Keywords:} Opinion dynamics, polarization and disagreement, crowd networks.

\noinhead{Paper type:} Literature review

\section{Introduction}
\label{sec:intro}
With the advance of communication and networking technology, the interactions among people are unprecedentedly enhanced.
People are free to express their own opinions, and interact with others through commenting, liking, retweeting in online social network platforms.
The increasing interactions sometimes result in fierce online debates \citep{durmus2019modeling,sridhar2015joint}.
There can be great opinion polarization and disagreement in the whole process, which might leads to online bullying \citep{squicciarini2015identification}.
In addition, some malicious people intend to spread misinformation in online social networks to sow discord in society, for example, during the 2016 presidential elections in the U.S. \citep{silva2020facebook} and the protest in Hong Kong \citep{zervopoulos2020hong}.
Such opinion polarization, disagreement and discord are harmful to the public security.
Therefore, it is of great importance to understand how people form opinions, evaluate the level of opinion polarization and disagreement, and prevent the harmful influence of such discord.

\begin{figure}[t!]
\centering
    \includegraphics[width=0.45\textwidth]{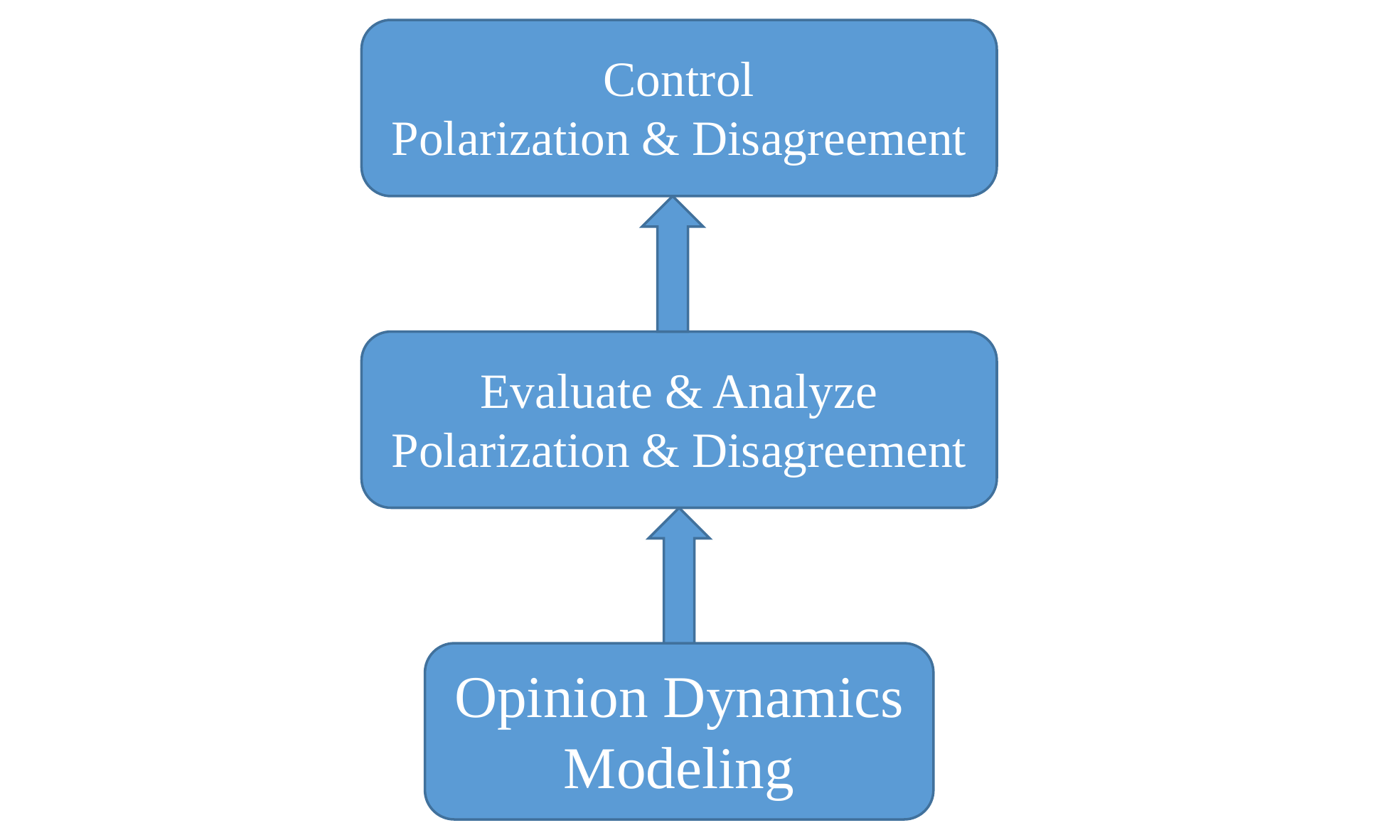}
    \caption{Three categories of works related to opinion polarization and disagreement and the relationships among these works.}
    \label{fig:work-cat}
\end{figure}

In literature, the works studying opinion polarization and disagreement can be classified into three categories: 1) opinion dynamics modeling; 2) evaluating and analyzing opinion polarization and disagreement; and 3) controlling opinion polarization and disagreement.
The relationship among the works of three categories is summarized in \reffig{fig:work-cat}.
Evaluating and analyzing opinion polarization and disagreement is based on the modeling of opinion dynamics.
With the opinion dynamic models and evaluations of polarization and disagreement, the works in the third category study how to control polarization and disagreement.

\subsection{Opinion Dynamics Modeling}
The opinion dynamics models can be classified into two categories based on whether opinions are discrete or continuous in the model.
In discrete models, the opinion value of individuals can either be binary, e.g., voting for Republicans or Democrats, or ordinal, e.g., the ratings of a movie (scores in $\{0, 1, 2, 3, 4, 5\}$).
However, in continuous models, the opinion values are real numbers, usually unified in range $[0,1]$ or $[-1, 1]$.
The lower and upper bounds represent the extreme opinion, e.g., completely support for Republicans or Democrats, respectively.
The opinion values in between can be interpreted as how close/far it is to/from the extreme opinion of upper/lower bound.
In the following, we briefly review the most basic discrete models and continuous models, respectively.

In discrete models, individuals are influenced by their neighbors, and update their opinions according to certain rules.
One seminal model is the voter model \citep{liggett2013stochastic}, where individuals randomly adopt one of his/her neighbors' opinion.
It will reach the opinion consensus state in voter model, where all individuals hold the same opinion.
\citeauthor{sood2005voter} find that both the first and second order of the degree distribution of the network influence the time to reach consensus \citep{sood2005voter}.

The foundation work of continuous opinion dynamics models is the DeGroot model \citep{degroot1974reaching}.
In this model, an individual updates opinion by averaging his/her neighbors' opinions.
By analyzing the equilibrium of such averaging process, this work shows that when the network is connected, it will reach opinion consensus state.
The Hegselmann-Krause (HK) opinion dynamics model is proposed based on the DeGroot model \citep{hegselmann2002opinion}.
It assumes that when the difference between two individuals' opinions is larger than a threshold, these two individuals will ignore each other's opinion when updating.
In this model, the opinions will finally converge to different clusters.
The opinions in the same cluster are the same, while those from different clusters are different.
In this case, the opinion polarization exists, that is, individuals hold different opinions at equilibrium state.
The work in \citep{castellano2009statistical} empirically finds that the number of opinion clusters is inversely proportional to the threshold value in the HK model.
One of the assumption in the HK model is that all individuals update their opinions at the same time.
\citeauthor{deffuant2000mixing} modified the model so that individuals update their opinions in an asynchronous manner \citep{deffuant2000mixing}.
The other important extension of the DeGroot model is the Fredkin-Johnsen (FJ) model \citep{friedkin1990social}.
In this model each individual is assigned with an \textit{internal opinion}, which represents his/her own belief on the topic.
When updating opinion, each individual also takes into account his/her internal opinion compared to the DeGroot model.
It is showed that, at the equilibrium of the FJ model, all individuals may hold different opinions, which is also opinion polarization state.
\citeauthor{bindel2015bad} explain the dynamics of the FJ model in an game-theoretic perspective.
Recently, the FJ model is widely adopted in the analysis of opinion polarization and disagreement \citep{musco2018minimizing,chen2018quantifying,matakos2017measuring} and opinion maximization \citep{gionis2013opinion,abebe2018opinion} due to its unique closed-from solution of equilibrium opinions.

\subsection{Contribution and Organization}
Opinion dynamics models have been studied in academia for decades.
The models we reviewed above are the most foundation works, and there are many variants of them in literatures including stubborn individuals \citep{wai2016active}, noise effect \citep{su2017noise}, and external sources \citep{majmudar2020voter}, etc.
There are some great works that systematically summarize and review the opinion dynamics models \citep{anderson2019recent,proskurnikov2017tutorial,proskurnikov2018tutorial,noorazar2020recent}.
However, little efforts have been made to review works on the evaluation and analysis of opinion polarization and disagreement, as well as the control of them.
Due to the recent harmful events that caused by discord, in this work, we aim at reviewing the recent advances on the study and control strategies of opinion polarization and disagreement.

Since most related works are based on the Fredkin-Johnsen (FJ) model, we first briefly introduce this model in \refsec{sec:prelim}.
Then, we review how polarization and disagreement are quantified in literatures and the relationships among them in \refsec{sec:analysis}.
Next, the works about controlling polarization and disagreement are reviewed in \refsec{sec:control}.
The conclusions and discussion is summarized in \refsec{sec:conclusion}.
\section{Preliminaries}
\label{sec:prelim}
In this section, we briefly introduce the FJ opinion dynamics model in \citep{bindel2015bad}
In the FJ model, there is a network with $n$ individual which can be modeled as a graph $G(V, E)$.
$V$ is the set of nodes representing individuals.
$E$ is the edge set, and for $(i, j) \in E$, the meaning is that individual $j$ can influence $i$.
The adjacency matrix $\mb{W}$ can be used to model the network whose entry-$i, j$ is $W_{ij}$, the influence weight of $j$ on $i$, if edge $(i, j) \in E$. Otherwise, $W_{ij} = 0$.
In literature, the network structure is usually assumed to be undirected, that is, $W_{ij} = W_{ji}$.
Another important concept about network is the graph laplacian $\mb{L}$.
Let $\mathbf{D}$ be the diagonal matrix $diag(d_1, \cdots, d_n)$, where $d_i = \sum_j W_{ij}$ is the degree of individual-$i$. Then, the definition of graph Laplacian is 
\begin{equation}
  \mb{L} = \mb{D} - \mb{W}.
\end{equation}
Both graph Laplacian $\mb{L}$ and the adjacency matrix $\mb{W}$ can characterize the network structure, because given one of them, the other one can be derived.

In the FJ model, individual-$i$ is assigned with an internal opinion $s_i$, which shows his/her beliefs about the discussed topic.
Let $\mb{s} = [s_1, \cdots, s_n]^T$ be the internal opinions of all individuals. 
The opinion formation process is divided into time step, and the internal opinions $\mb{s}$, is assumed to be a constant.
The express opinion of individuals at time step $t$ is $\mb{z}(t) = [z_1(t), \cdots, z_n(t)]^T$, where entry $z_i(t)$ is the express opinion of individual-$i$ at time $t$.
Let $N_i$ be the set of individual-$i$'s neighbors.
In this model, individual-$i$ updates his/her express opinion as
\begin{equation}
  z_i(t+1) = \frac{s_i + \sum_{j\in N_i} W_{ij} z_j(t)}{1 + \sum_{j\in N_i} W_{ij}}.
\end{equation}
That is, the express opinion for individual-$i$ at the next time step $t+1$ is the weighted average of his/her internal opinion and the express opinions of his/her neighbors at this time step $t$.

The opinion formation process evolves and reach the equilibrium, where no individuals' opinions change any more.
Let $\mb{z}$ be the equilibrium opinions.
According to the definition of equilibrium, we have
\begin{equation}
  \mb{z} = (\mb{L} + \mb{I})^{-1} \cdot \mb{s},
  \label{eqn:equilibrium}
\end{equation}
where $\mb{I}$ is identity matrix with size $n\times n$ and $\mb{L}$ is the graph Laplacian of the network.
From the \refeq{eqn:equilibrium}, we know that the equilibrium opinions depend on the internal opinions $\mb{s}$ and the graph structure $\mb{L}$ (or $\mb{W}$).
In addition, the choice of initial express opinions $\mb{z}(0)$ does not influence the equilibrium.
\citeauthor{friedkin1990social} show that the opinions at equilibrium can be different, that is, there can be opinion polarization at equilibrium.
In the next, we review works that evaluate and analyze opinion polarization and disagreement based on the FJ model.
The notations in this review is summarized in \reftab{tab:notations}.

\begin{table}
  \renewcommand\arraystretch{1.2}
  \centering
  \caption{Notations}
  \scalebox{0.9}{
  \begin{tabular}{c|c|c}
    \hline
    \multicolumn{2}{c|}{Symbol}  &   Meaning\\
    \hline
    \multirow{4}*{\makecell[c]{Opinion\\Related}} & $\mb{s}$, $s_i$  &   All internal opinions, internal opinion of individual-$i$\\
    ~ & $\mb{z}(t)$, $z_i(t)$   &   All express opinions at time $t$, express opinion of individual-$i$ at time $t$\\
    ~ & $\mb{z}$, $z_i$  &  All equilibrium opinions, equilibrium opinion of individual-$i$\\
    ~ & $\bar{\mb{s}}$, $\bar{\mb{z}}$  &  Mean-centered internal opinions and equilibrium opinion\\
    \hline
    \multirow{3}*{\makecell[c]{Network\\Related}} & $\mb{W}$ & The adjacency matrix of the network\\
    ~ & $\mb{L}$ & The graph Laplacian of the network\\
    ~ & $d_i$   & The degree of individual-$i$\\
    \hline
    \multirow{4}*{\makecell[c]{Metric\\Related}} & $\mathcal{P}$   &  Opinion polarization\\
    ~ & $\mathcal{D}$   &  Opinion disagreement\\
    ~ & $PDI(\mu)$    &   polarization-disagreement index with tradeoff factor $\mu$\\
    ~ & $PDI$         &   PDI with tradeoff factor 1\\
    \hline
  \end{tabular}}
  \label{tab:notations}
\end{table}

\section{Evaluation and Analysis of Polarization and Disagreement}
\label{sec:analysis}
In this section, we first review how opinion polarization, disagreement, and other related metrics are defined and evaluated in literature.
Then, we summarize the analyses about opinion polarization and disagreement.

\subsection{Quantifying Polarization and Disagreement}
There are some works analyze the opinion polarization and disagreement based on the Fredkin-Johnsen (FJ) opinion dynamics model \citep{chen2018quantifying,musco2018minimizing,dandekar2013biased}.
Let $\mb{z}$ be the opinions at equilibrium in the FJ model, and
\begin{equation}
  \bar{z} = \frac{1}{n} \mb{1}^T \mb{z}
\end{equation}
be the mean of equilibrium opinions.
The mean-centered equilibrium opinions are
\begin{equation}
  \bar{\mb{z}} = \mb{z} - \bar{z}\cdot \mb{1} = (\mb{I} - \frac{1}{n} \mb{1}\mb{1}^T)\cdot \bar{\mb{z}}.
\end{equation}
Similarly, the mean-centered internal opinions $\bar{\mb{s}}$ is defined as
\begin{equation}
  \bar{\mb{s}} = \mb{s} - \bar{s}\cdot \mb{1} = (\mb{I} - \frac{1}{n} \mb{1}\mb{1}^T)\cdot \bar{\mb{s}}, \text{ where}
\end{equation}
\begin{equation}
  \bar{s} = \frac{1}{n} \mb{1}^T \mb{s}
\end{equation}
is the average value of internal opinions.
It is shown in \citep{musco2018minimizing} that
\begin{equation}
  \bar{\mb{z}} = (\mb{L} + \mb{I})^{-1} \cdot \bar{\mb{s}}.
\end{equation}

\noinhead{Polarization}. The polarization is defined as
\begin{equation}
  \begin{aligned}
    \mathcal{P} &= \sum_{i} (z_i - \bar{z})^2 = \bar{\mb{z}}^T\bar{\mb{z}} = \bar{\mb{s}}^T (\mb{L} + \mb{I})^{-2} \bar{\mb{s}}\\
    &= \mb{s}^T (\mb{L} + \mb{I})^{-1} (\mb{I} - \frac{1}{n}\mb{1}\mb{1}^T) (\mb{L} + \mb{I})^{-1} \mb{s},
  \end{aligned}
\end{equation}
which is the variance of equilibrium opinions.
From the definition, we can see that polarization measure how equilibrium opinions deviate from the average \citep{musco2018minimizing}.

\noinhead{Disagreement}. Different from polarization, disagreement quantifies the extent to which the express opinions of neighbors are in different with each other \citep{chen2018quantifying}.
First, the local disagreement on edge $(i, j)\in E$ is defined as
\begin{equation}
  \begin{aligned}
    d(i,j) &= W_{ij}\cdot (z_i - z_j)^2\\
    &= W_{ij} \cdot ((z_i - \bar{z}) - (z_j - \bar{z}))^2\\
    &= W_{ij} \cdot(\bar{z}_i - \bar{z}_j)^2.
  \end{aligned}
\end{equation}
The above equation also shows that local disagreement on edge $(i,j)$ can be calculated through either equilibrium opinions $\mb{z}$ or the mean-centered ones $\bar{\mb{z}}$.
Then, the disagreement of the whole network is defined as the sum of all local disagreement on edges, that is,
\begin{equation}
  \mathcal{D} = \sum_{(i, j) \in E} d(i, j).
\end{equation}
It is shown in \citep{musco2018minimizing} that
\begin{equation}
  \begin{aligned}
    \mathcal{D} &= \bar{\mb{z}}^T \mb{L} \bar{\mb{z}}= \bar{\mb{s}}^T (\mb{L}+\mb{I})^{-1} \mb{L} (\mb{L} + \mb{I})^{-1} \bar{\mb{s}}\\
    &= \mb{s}^T (\mb{L}+\mb{I})^{-1} \mb{L} (\mb{L} + \mb{I})^{-1} \mb{s}.
  \end{aligned}
\end{equation}

\noinhead{Polarization-Disagreement Index (PDI)}.
This metric combines both opinion polarization and disagreement in a weighted average manner, that is,
\begin{equation}
  PDI(\mu)= \mathcal{P} + \mu\cdot\mathcal{D},
\end{equation}
where the hyperparameter $\mu$ represents the importance of opinion disagreement to PDI comparing with opinion polarization.
In this review, we denote $PDI$ as $PDI(1)$, where opinion polarization and disagreement contribute equally.
For $PDI$, we have
\begin{equation}
  \begin{aligned}
    PDI &= \mathcal{P} + \mathcal{D} = \bar{\mb{z}}^T (\mb{L} + \mb{I}) \bar{\mb{z}}= \bar{\mb{s}}^T (\mb{L} + \mb{I})^{-1} \bar{\mb{s}}\\
    &= \mb{s}^T (\mb{I} - \frac{1}{n}\mb{1}\mb{1}^T) (\mb{L} + \mb{I})^{-1} (\mb{I} - \frac{1}{n}\mb{1}\mb{1}^T) \mb{s}.
  \end{aligned}
\end{equation}

\begin{table}
  \renewcommand\arraystretch{1.2}
  \centering
  \caption{Definition of polarization and disagreement.}
  \scalebox{0.9}{
  \begin{tabular}{|c|c|c|c|}
    \hline
    Items & Through $\bar{\mb{z}}$ & Through $\bar{\mb{s}}$ & Through $\mb{s}$\\
    \hline
    Polarization: $\mathcal{P}$ & $\bar{\mb{z}}^T\bar{\mb{z}}$  & $\bar{\mb{s}}^T (\mb{L} + \mb{I})^{-2} \bar{\mb{s}}$ & $\mb{s}^T (\mb{L} + \mb{I})^{-1} (\mb{I} - \frac{1}{n}\mb{1}\mb{1}^T) (\mb{L} + \mb{I})^{-1} \mb{s}$ \\
    Disagreement: $\mathcal{D}$ & $\bar{\mb{z}}^T \mb{L} \bar{\mb{z}}$ & $\bar{\mb{s}}^T (\mb{L}+\mb{I})^{-1} \mb{L} (\mb{L} + \mb{I})^{-1} \bar{\mb{s}}$ & $\mb{s}^T (\mb{L}+\mb{I})^{-1} \mb{L} (\mb{L} + \mb{I})^{-1} \mb{s}$ \\
    PDI: $\mathcal{P} + \mathcal{D}$ & $\bar{\mb{z}}^T(\mb{L} + \mb{I})\bar{\mb{z}}$ & $\bar{\mb{s}}^T (\mb{L} + \mb{I})^{-1} \bar{\mb{s}}$ & $\mb{s}^T (\mb{I} - \frac{1}{n}\mb{1}\mb{1}^T) (\mb{L} + \mb{I})^{-1} (\mb{I} - \frac{1}{n}\mb{1}\mb{1}^T) \mb{s}$ \\
    \hline
  \end{tabular}}
  \label{tab:pol-dis}
\end{table}

\noinhead{Remarks on Polarization and Disagreement}. The polarization, disagreement related metrics defined above are summarized in \reftab{tab:pol-dis}.
From the definition, we can see that they are all quadratic forms of internal opinions $\mb{s}$ (or mean-centered express opinion $\bar{\mb{z}}$ and mean-centered internal opinion $\bar{\mb{s}}$).
In addition, as shown in \citep{gaitonde2020adversarial}, these three quadratic forms are all positive semi-definite, and thus, they are all convex functions with respect to $\mb{s}$ (or $\bar{\mb{s}}$ and $\bar{\mb{z}}$).

\subsection{Analysis of Polarization and Disagreement}
\begin{figure}[t!]
\centering
\subfigure[Example 1]{
  \includegraphics[width=0.45\textwidth]{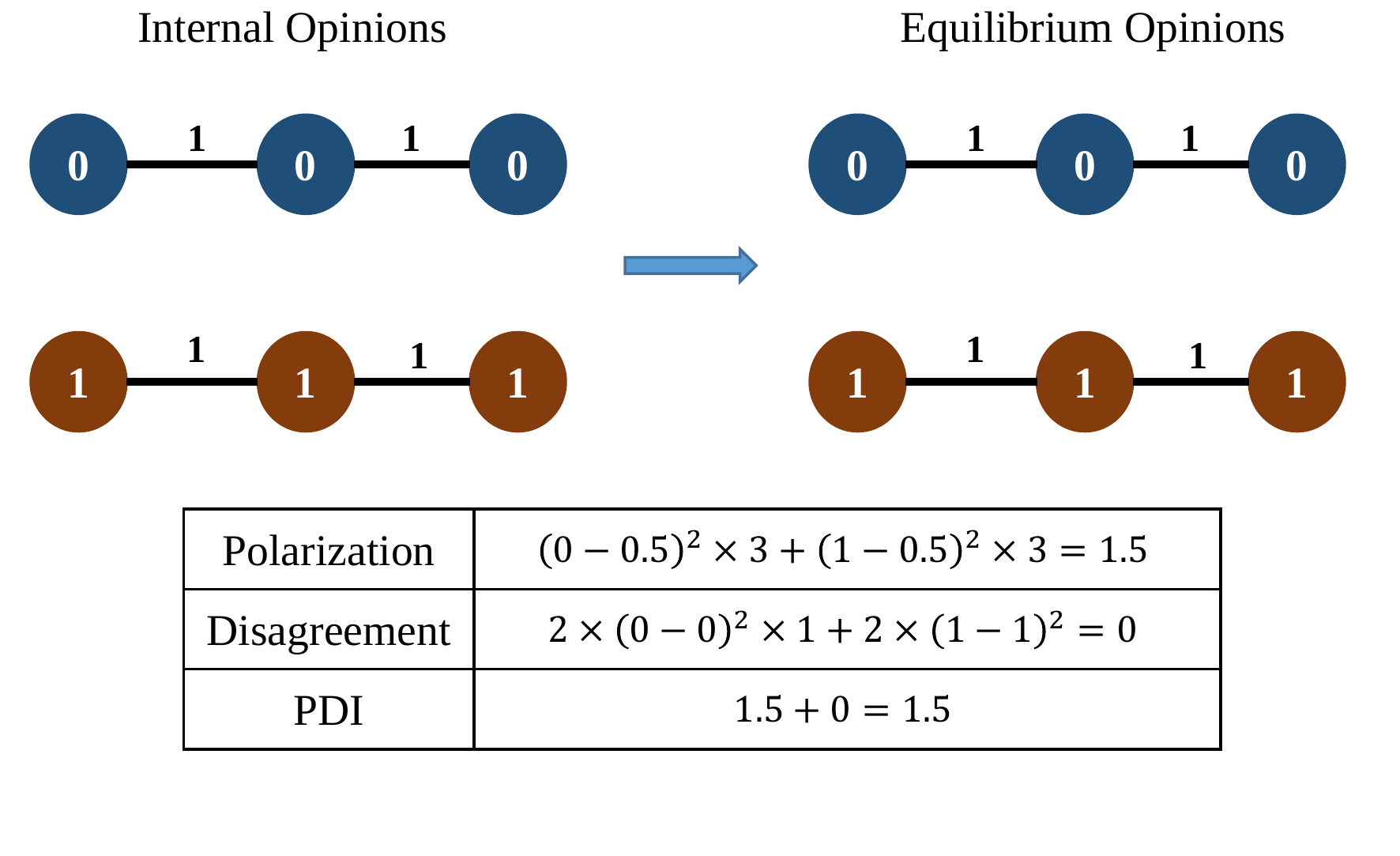}
  \label{fig:two-egs-1}
}
\quad
\subfigure[Example 2]{
  \includegraphics[width=0.45\textwidth]{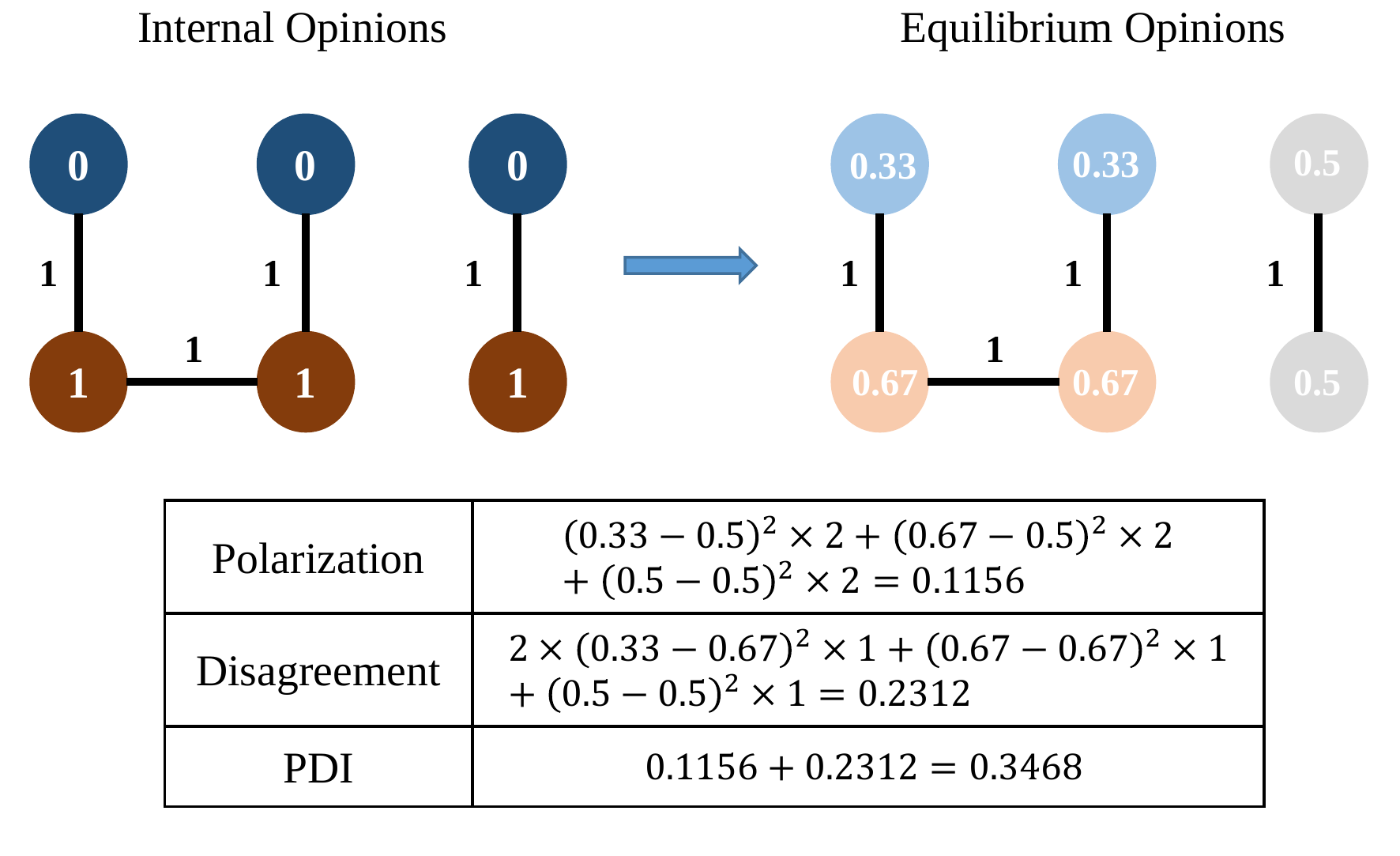}
  \label{fig:two-egs-2}
}
\caption{Two examples for polarization and disagreement.}
\label{fig:two-egs}
\end{figure}
Since both opinion polarization and opinion disagreement are harmful to the public security, both of them are expected to be weakened.
However, this is hard to achieve \citep{musco2018minimizing}.
Consider the examples in \reffig{fig:two-egs}, there are both six individuals in the network.
Three of them have internal opinion $0$, while the other three of them have opinion $1$.
Both networks have four edges with weight $1$.
We can see that the polarization and disagreement at the equilibrium are different in the two networks.
The network in \reffig{fig:two-egs-1} has lower disagreement and higher polarization, while that in \reffig{fig:two-egs-2} has higher disagreement and lower polarization.
This is because that the network in \reffig{fig:two-egs-1} only connects individuals with the same internal opinion.
This forms ``echo chamber'' \citep{jamieson2008echo} where individuals only interact with those who have the similar opinions, and their interactions further enhance their opinions.
While the network in \reffig{fig:two-egs-2} connects individuals with different internal opinions.
According to the FJ model, the individuals with different opinions influence with each other, and their express opinions get closer to others.
Therefore, the polarization in this network is small.

\begin{algorithm}[!t]
\SetAlgoLined
  \KwIn{Initial graph Laplacian $\hat{\mb{L}}$\;
  Initial internal opinion $\hat{\mb{s}}$\;
  Repeated round number \texttt{ROUND}.}
  \KwOut{Expressed opinions after \texttt{ROUND} $\mb{z}$}
    \For{$r = 1, \cdots, $ \texttt{ROUND}}{
      $\mb{z} \leftarrow (\mb{L} + \mb{I})^{-1} \cdot \mb{s}$ \tcp*{\texttt{Opinion updating}}
      Solve \refeq{eqn:net-admin} and obtain new network adjacency matrix $\mb{W}$ \tcp*{Weight adjusting.}
      Update graph Laplacian $\mb{L}$ with $\mb{W}$.
    }
    \caption{Opinion dynamics with network administrator.}
    \label{alg:op-dyn-net-admin}
\end{algorithm}
\citeauthor{chitra2020analyzing} further analyze opinion polarization and disagreement in real online social networks.
They introduce \emph{network administrator} into the FJ opinion dynamics model, whose function is to increase individual engagement via personalized filtering, or showing individuals content that they are more likely to agree with.
This corresponds to reducing opinion disagreement by adjusting edge weights of the graph in the FJ model (e.g. individuals see more content from the others with similar opinions).
Their proposed opinion dynamics with network administrator is shown in \refalg{alg:op-dyn-net-admin}.
Specifically, the dynamics with network administrator includes multiple rounds.
Suppose that the initial graph adjacency matrix is $\widehat{\mb{W}}$ and the internal opinions are $\mb{s}$.
In each round, all individuals first update their opinions according to the FJ model and reach the equilibrium $\mb{z}$.
Then, the network administrator adjusts the network structure to minimize disagreement based on the equilibrium opinions, that is,
\begin{equation}
  \begin{array}{lc}
    \min_{\mb{W}} & \mathcal{D}=\bar{\mb{z}}^T\mb{L}\bar{\mb{z}}\\
    s.t.  & \bar{\mb{z}} = (\mb{I} - \frac{1}{n}\mb{1}\mb{1}^T)\cdot \mb{z}\\
          & \mb{L} \text{ is graph Laplacian of } \mb{W}\\
          & \Vert \mb{W} - \widehat{\mb{W}} \Vert_F \le \epsilon\cdot \Vert\widehat{\mb{W}}\Vert_F\\
          & \sum_j W_{ij} = \sum_j \widehat{W}_{ij}.
  \end{array}
  \label{eqn:net-admin}
\end{equation}
The last two constraints in the above optimization problem ensure that the total change of weights is bounded, and the total weights of each individual remain unchanged.
The whole process repeated until it converges.

\citeauthor{chitra2020analyzing} further validate the proposed model with Twitter and Reddit data in \citep{de2014learning}.
They showed that as $\epsilon$ in \refeq{eqn:net-admin} increase, that is, network administrator can adjust more weights of network, the polarization increases surprisingly fast while disagreement shrinks.
This observation further validates that there is a tradeoff between opinion polarization and disagreement.
In addition, the network administrator in this work acts as recommender systems in online social networks, and their recommender behavior (exposing individuals with similar opinion to each other) can cause ``filter bubble'' effect \citep{pariser2011filter}, which have been blamed for causing severe opinion polarization in social science and psychology \citep{bakshy2015exposure,garimella2018political,stroud2010polarization}.

\section{Control of Polarization and Disagreement}
\label{sec:control}
Based on the analyses of opinion polarization and disagreement, a key problem is how to control them.
With the definition of metrics related to polarization and disagreement in \reftab{tab:pol-dis}, we can see that controlling polarization and disagreement can be done by tuning the network structure $\mb{L}$ (or $\mb{W}$) or the internal opinions $\mb{s}$.

\subsection{Control over Network structure}
As the example in \reffig{fig:two-egs}, the network structure has large impact on polarization and disagreement.
Since the graph Laplacian $\mb{L}$ shows how individuals are influenced by each other, tuning $\mb{L}$ can be explained as interfering the interactions among individuals.
\citeauthor{musco2018minimizing} first considered to minimize the polarization-disagreement index with the graph Laplacian.
They assumed that the total weight of the network is a constant $m$, and the problem was formulated as
\begin{equation}
  \begin{array}{lc}
  \min_{\mb{L}}  & PDI = \bar{\mb{s}}^T (\mb{L} + \mb{I})^{-1} \bar{\mb{s}}\\
  s.t.      & Tr(L) = m,
  \end{array}
  \label{eqn:musco-2018-net}
\end{equation}
Where the constraint means that the total weights of the network remains a constant $m$.
Furthermore, \citeauthor{musco2018minimizing} show that \refeq{eqn:musco-2018-net} is a convex optimization problem with respect to $\mb{L}$.
However, if the objective function in \refeq{eqn:musco-2018-net} is $PDI(\mu)$ where $\mu\neq 1$, the convexity does not hold anymore.

Although \refeq{eqn:musco-2018-net} is convex, there are $n\times n$ (the size of $\mb{L}$) variables to be decided, which consumes large memory and time when solving.
In addition, if the solution corresponds to a dense network, that is, a lot of entries of $\mb{L}$ is not zero, this means that the interactions between any two individuals need to be adjusted precisely.
This is infeasible in reality under the constraints of limited resources and time.
To overcome the above two issues, \citeauthor{musco2018minimizing} implement the sparse algorithm in \citep{spielman2011graph,spielman2011spectral,spielman2014nearly} to effectively solve \refeq{eqn:musco-2018-net} and obtain a suboptimal solution which has much less edges.
According to their experiments on synthetic networks, the suboptimal solution have only about $\frac{1}{7}$ edges compared to the optimal solution to \refeq{eqn:musco-2018-net}, while the gap between $PDI$ calculated by suboptimal solution and the optimal solution is negligible.

\citeauthor{chen2018quantifying} argue that it is expected to minimize polarization and disagreement of a certain topic by tuning network structure, before this topic begins to be discussed.
However, the metrics in \reftab{tab:pol-dis} are all related to internal opinions, which is hard to obtain before the topic begins.
To achieve this goal, \citeauthor{chen2018quantifying} regard the mean-centered internal opinions $\bar{\mb{s}}$ as a random vector, and define the \emph{Average-case Conflict Risk (ACR)} of metrics in \reftab{tab:pol-dis}.
Note that all metrics in \reftab{tab:pol-dis} can be expressed in the form of $\bar{\mb{s}}^T \mb{M}_* \bar{\mb{s}}$, where $\mb{M}_*$ is the positive semi-definite matrix of metric $*$, that is, $\mb{M}_{\mathcal{P}} = (\mb{L}+\mb{I})^{-2}$ for polarization, $\mb{M}_{\mathcal{D}} = (\mb{L}+\mb{I})^{-1} \mb{L} (\mb{L} + \mb{I})^{-1}$ for disagreement, and $\mb{M}_{PDI} = (\mb{L}+\mb{I})^{-1}$ for polarization-disagreement index (PDI).
The ACR assumes that all mean-centered internal opinions are independent and they follow uniform distribution in $[-1, 1]$.
Therefore, $E[\bar{\mb{s}}\bar{\mb{s}}^T] = \mb{I}$.
The ACR is defined as the mean of a metric, that is,
\begin{equation}
  ACR_* = E[\bar{\mb{s}}^T \mb{M}_* \bar{\mb{s}}] = E[Tr(\bar{\mb{s}}\bar{\mb{s}}^T \mb{M}_* )] = Tr(E[\bar{\mb{s}}\bar{\mb{s}}^T]\mb{M}_* ) = Tr(\mb{M}_* ).
\end{equation}
Furthermore, \citeauthor{chen2018quantifying} formulate the problem to minimize $ACR_*$ by controlling the network structure as
\begin{equation}
    \begin{array}{lc}
        \min_{\mb{W}}    &   ACR_{*} = Tr(\mb{M}_*)\\
        s.t.        &   0 \le \mb{W} \le 1\\
                    &   \Vert \mb{W} - \widehat{\mb{W}} \Vert_1 \le k,
    \end{array}
    \label{eqn:acr}
\end{equation}
where the first constraint means that each edge weight is in range $[0,1]$.
The norm $\Vert \cdot \Vert_1$ in the second constraint is the entry-wise one-norm.
Therefore, the second constraint shows that the difference between $\mb{W}$ and a known adjacency matrix $\widehat{\mb{W}}$ is bounded by $k$.
It is shown in \citep{chen2018quantifying} that only $ACR_{PDI}$ is convex.
\citeauthor{chen2018quantifying} empirically find that the complete network where all edges' weights are $1$ can both minimize $ACR_{PDI}$ and $ACR_{\mathcal{P}}$ in \refeq{eqn:acr}.
However, for the network that can minimize $ACR_{\mathcal{D}}$, it contains sets of disconnected subgraphs which are cliques, trees, and chains.
\citeauthor{chen2018quantifying} argue that the disconnected cliques, trees and chains network structure seem to correspond with common management structures in companies: a flat organization corresponds to a clique, while a hierarchical organization corresponds to a tree. 
In the perspective of companies' interests, it is often assumed to reduce disagreement.
Therefore, the learned network structure can provide guidance for companies' team construction.

\subsection{Control over Internal Opinion}
In literature, there are some works assume that the network structure $\mb{W}$ or ($\mb{L}$) is known, and try to manipulate individuals' internal opinions in order to control the polarization and disagreement.

\citeauthor{musco2018minimizing} propose to control internal opinions in order to minimize the polarization-disagreement index.
We can show from the definition of polarization and disagreement that if all individuals hold the same internal opinion, for example $\mb{s} = \mb{0}$, both polarization $\mathcal{P}$ and disagreement $\mathcal{D}$ reach their minima $0$.
This trivial solution exists because there is no constraints on the internal opinions.
In \citep{musco2018minimizing}, the author propose the problem that given an internal opinion $\mb{s}$, how to change the internal opinion constrained by a constraint $\alpha$ so that the polarization-disagreement index is minimized?
The mathematical formulation is
\begin{equation}
    \begin{array}{lc}
        \min_{\mb{d}}    &   (\mb{s} - \mb{d})^T (\mb{I} - \frac{1}{n}\mb{1}\mb{1}^T) (\mb{L} + \mb{I})^{-1} (\mb{I} - \frac{1}{n}\mb{1}\mb{1}^T) (\mb{s} - \mb{d})\\
        s.t.        &   \mb{0} \le \mb{d} \le \mb{s}\\
                    &   \mb{1}^T \mb{d} \le \alpha,
    \end{array}
\end{equation}
where $\mb{d}$ is the changing vector to the internal opinions.
The above optimization problem is convex (specifically semi-definite programming), and can be solved efficiently with techniques, for example, interior point method, in polynomial time \citep{boyd2004convex}.
\citeauthor{musco2018minimizing} run the above optimization on synthetic network with the power-law degree distribution \citep{newman2005power}, and find that the internal opinions which is large tend to be reduced most.
In addition, the authors examine the equilibrium opinion after optimization and find that all opinions at equilibrium are close to each other.
If the internal opinions follows the power-law distribution before optimization, the equilibrium opinions after optimization tend to be centered around $0$.
However, if the internal opinion follows the uniform distribution before optimization, the equilibrium opinions after optimization tend to be centered around $0.5$. 

The above work assumes that all individuals' innate opinions can be manipulated, which is hard to achieve in reality.
From the perspective of adversary, \citeauthor{chen2020network} aim at maximizing polarization and disagreement by only controlling a few individuals' internal opinion.
The individuals who are controlled by adversary is called \emph{target individual}.
Given the internal opinions $\mb{s}$, let $\mathcal{P}(\mb{s})$ and $\mathcal{D}(\mb{s})$ be the polarization and disagreement, respectively, according to \reftab{tab:pol-dis}.
Then, with the known internal opinions $\hat{\mb{s}}$ and the graph Laplacian of the network, the maximization problem of adversary is formulated as
\begin{equation}
  \begin{array}{lc}
    \max_{\mb{s}} & \mathcal{P}(\mb{s}) \\
      s.t.        & \Vert \mb{s} - \hat{\mb{s}}\Vert_0 = k\\
                  & 0 \le \mb{s} \le 1, \text{   and}
  \end{array}
  \label{eqn:pol-dis-cons}
\end{equation}
\begin{equation}
  \begin{array}{lc}
    \max_{\mb{s}} & \mathcal{D}(\mb{s})\\
      s.t.        & \Vert \mb{s} - \hat{\mb{s}}\Vert_0 = k\\
                  & 0 \le \mb{s} \le 1.
  \end{array}
  \label{eqn:dis-dis-cons}
\end{equation}
The objectives of the two optimization problems are maximizing opinion polarization and disagreement, respectively.
This is because from the perspective of the adversaries, they want the society to be in chaos, and the opinion polarization and disagreement to be large.
The constraints $\Vert \mb{s} - \hat{\mb{s}}\Vert_0 = k$ limits the resources of the adversaries, which means that only $k$ of individuals' internal opinions can be controlled.
By solving these two problems, the adversaries could find $k$ targeted individuals, and change their internal opinions correspondingly by, for example, persuasion.
\citeauthor{chen2020network} derive the following inequalities:
\begin{align}
  \mathcal{P}(\mb{s}_{\mathcal{P}}) &\le  \mathcal{P}(\hat{\mb{s}}) + 3 k \text{   and} \\
  \mathcal{D}(\mb{s}_{\mathcal{D}}) &\le  \mathcal{D}(\hat{\mb{s}}) + 8 d_{max} k,
\end{align}
where $\mb{s}_{\mathcal{P}}$ and $\mb{s}_{\mathcal{D}}$ are the optimal solution to \refeq{eqn:pol-dis-cons} and \refeq{eqn:dis-dis-cons}, respectively, and $d_{max}$ is the largest degree of the given network.
The above two bounds show that both the increase of polarization and disagreement is bounded linearly by $k$.

\begin{figure}[t!]
  \centering
  \subfigure[Polarization]{
    \includegraphics[width=0.45\textwidth]{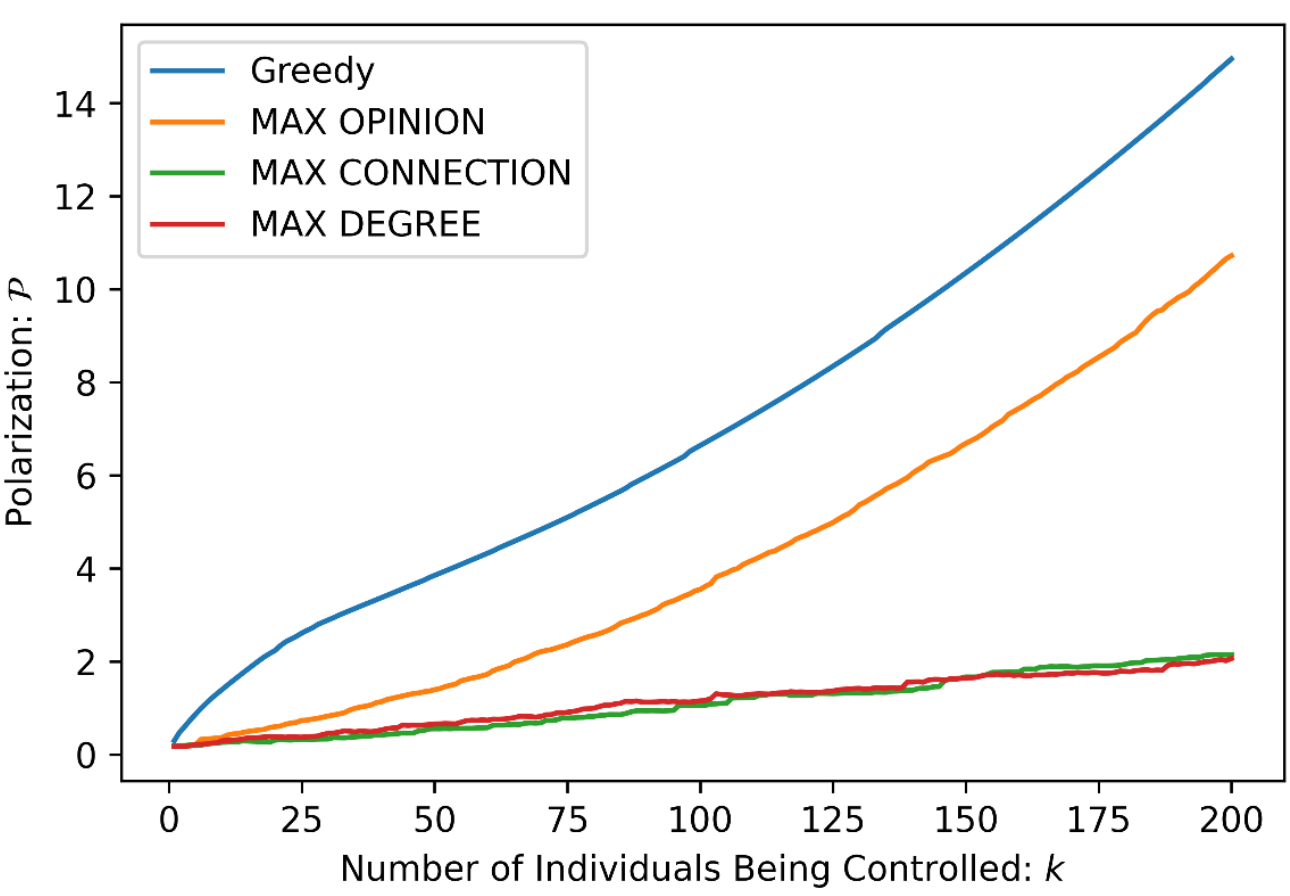}
    \label{fig:dis-compare-pol}
  }
  \quad
  \subfigure[Disagreement]{
    \includegraphics[width=0.45\textwidth]{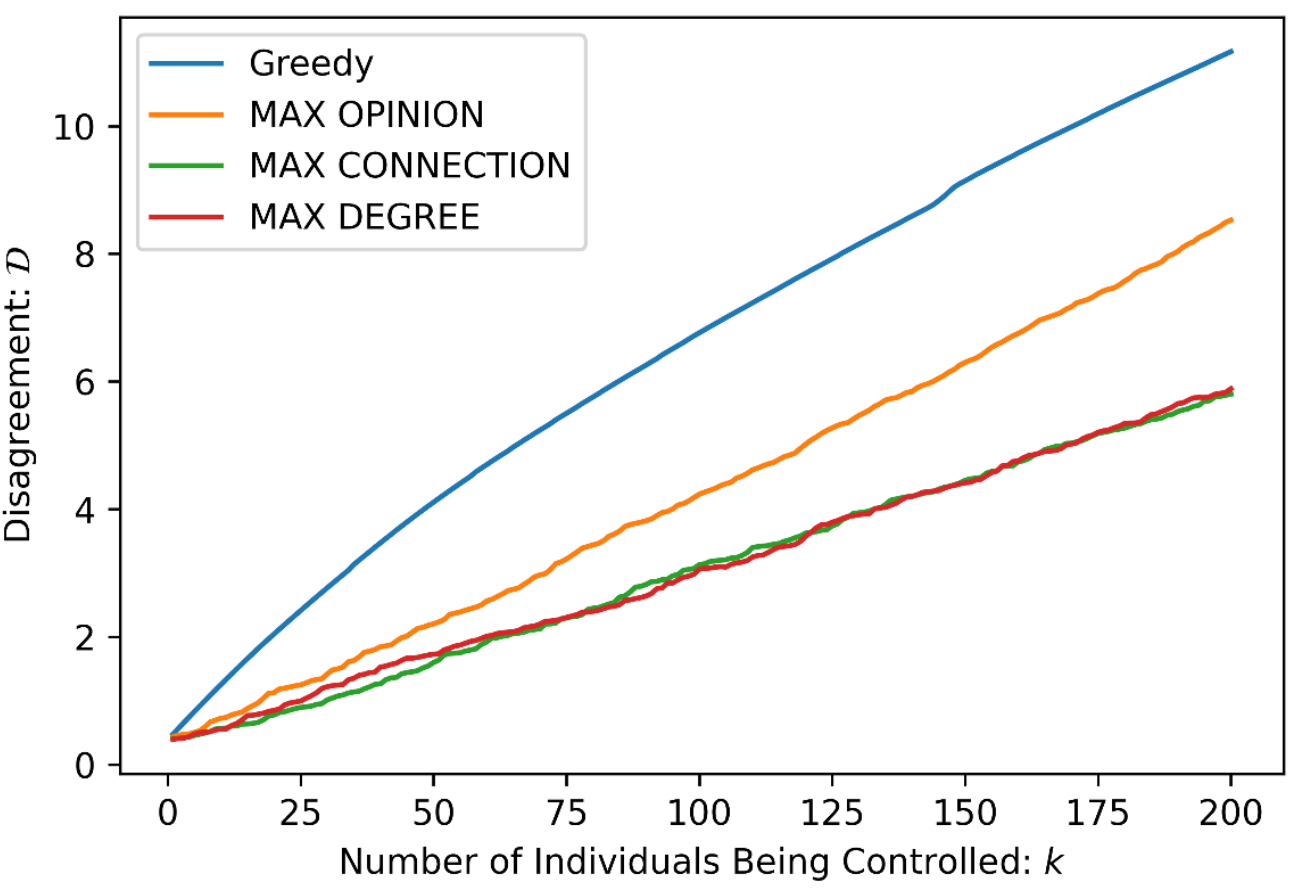}
    \label{fig:dis-compare-dis}
  }
  \caption{Polarization and disagreement after manipulating $k$ individuals' internal opinions.}
  \label{fig:dis-compare}
  \end{figure}

\begin{algorithm}[!t]
\SetAlgoLined
  \KwIn{Initial graph laplacian $\hat{\mb{L}}$\;
  Initial internal opinion $\hat{\mb{s}}$\;
  Number of target individuals $k$.}
  \KwOut{The set of target individuals $\Omega$\;
  The manipulated internal opinion $\mb{s}$.}
  $\mb{s} \leftarrow \hat{\mb{s}}$ and $\Omega \leftarrow \emptyset$ \tcp*[r]{Initialization.}
  \For{$i = 1, \cdots, k$}{
    \tcp{Find only one individual that can maximize $\mathcal{P}$ (or $\mathcal{D}$) in each iteration.}
    $maxVal \leftarrow 0$ \tcp*[r]{The maxima of $\mathcal{P}$ (or $\mathcal{D}$)}
    $index \leftarrow 0$  \tcp*[r]{The individual that maximize $\mathcal{P}$ (or $\mathcal{D}$)}
    $setVal \leftarrow 0$ \tcp*[r]{The internal opinion value set to individual $index$}
    \For{$j = 1, \cdots, n$}{
      \If{$j \notin \Omega$}{
        $\mb{s}^\prime = \mb{s}$\;
        \tcp{Enumerate two extreme opinion for individual $j$.}
        Set $s^\prime_j$ to $0$, obtain $\mb{s}^\prime_0$, calculate $\mathcal{P}(\mb{s}^\prime_0)$ (or $\mathcal{D}(\mb{s}^\prime_0)$)\;
        \If{$\mathcal{P}(\mb{s}^\prime_0) \ge maxVal$ \tcp*[r]{$\mathcal{D}(\mb{s}^\prime_0) \ge maxVal$ for $\mathcal{D}$}}{
          $maxVal \leftarrow \mathcal{P}(\mb{s}^\prime_0$), $index \leftarrow j$, and $setVal \leftarrow 0$\;
          \tcp{$maxVal \leftarrow \mathcal{D}(\mb{s}^\prime_0$) for $\mathcal{D}$}
        }
        Set $s^\prime_j$ to $1$, obtain $\mb{s}^\prime_1$, calculate $\mathcal{P}(\mb{s}^\prime_1)$ (or $\mathcal{D}(\mb{s}^\prime_1)$)\;
        \If{$\mathcal{P}(\mb{s}^\prime_1) \ge maxVal$ \tcp*[r]{$\mathcal{D}(\mb{s}^\prime_1) \ge maxVal$ for $\mathcal{D}$}}{
          $maxVal \leftarrow \mathcal{P}(\mb{s}^\prime_1$), $index \leftarrow j$, and $setVal \leftarrow 1$\;
          \tcp{$maxVal \leftarrow \mathcal{D}(\mb{s}^\prime_0$) for $\mathcal{D}$}
        }
      }
    }
    Change the $index$ entry of $\mb{s}$ to $setVal$ \tcp*[r]{Update $\mb{s}$}
    $\Omega \leftarrow \Omega \cup \{index\}$ \tcp*[r]{Update $\Omega$}
  }
  \caption{Greedy algorithm for maximizing polarization or disagreement.}
  \label{alg:greedy}
\end{algorithm}
Due to the convexity of polarization $\mathcal{P}$, disagreement $\mathcal{D}$, \citeauthor{chen2020network} first show that any internal opinion of target individual must be set to the extreme opinion, that is, either $0$ or $1$.
There are $\begin{pmatrix} n \\ k \end{pmatrix}$ cases to choose $k$ out of $n$ individuals.
Setting each of chosen individual to $0$ or $1$ requires $2^k$ enumerations.
Therefore, it requires $\begin{pmatrix} n \\ k \end{pmatrix}\cdot 2^k$ enumerations to decide the optimal solution to \refeq{eqn:pol-dis-cons} and \refeq{eqn:dis-dis-cons}, and it is infeasible to use such brute force enumeration method.
\citeauthor{chen2020network} propose to use the hill-climbing greedy algorithm in \refalg{alg:greedy} to solve the above problems \citep{domingos2001mining,richardson2002mining,kempe2003maximizing}.
The greedy algorithm iteratively finds $k$ target individuals, that is, it only find one individual that can maximize polarization or disagreement and the internal opinion of him/her in each iteration.
This process continues until $k$ individuals are found.
Compared to the brute force algorithm, it only need $k\times n \times 2$ enumerations, which significantly reduces the computational complexity.
In addition, \citeauthor{chen2020network} also choose the following heuristic methods to decide which $k$ individuals to choose and how their internal opinions should be set.
\begin{itemize}
  \item \texttt{MEAN OPINION}. Choose the $k$ individuals who have the internal opinions that are the closest to the mean internal opinion.
  \item \texttt{MAX CONNECTION}. Choose the $k$ individuals who have the most connections with other individuals, that is, the corresponding rows in the adjacency matrix which have the most non zero entries.
  \item \texttt{MAX DEGREE}. Choose the $k$ individuals who have the largest degree.  
\end{itemize}
With the chosen $k$ individuals, their internal opinions are set to either $0$ or $1$, respectively, so that polarization or disagreement is maximized.
The performances of greedy algorithm and heuristic algorithms are tested with Twitter data in \citep{de2014learning}, and shown in \reffig{fig:dis-compare} \footnotetext{The results are implemented from https://github.com/mayeechen/network-disruption}.
We can see that the greedy algorithm is superior to other heuristic methods on both maximizing polarization and disagreement.
The \texttt{MEAN OPINION} algorithm performs the best among heuristic methods.
One possible reason is that this algorithm intensionally separate internal opinions which are originally close to each other to different extremes (either $0$ or $1$).
The original ``friends'' who have similar opinions and interests are provoked by adversary, and their friendships would be broken.
In this way, the polarization and disagreement can be greatly enhanced.

\section{Conclusion and Discussion}
\label{sec:conclusion}
In this paper, we review works of evaluating and controlling opinion polarization and disagreement.
Based on the Fredkin-Johnsen opinion dynamics model, polarization and disagreement are defined in literature.
Polarization shows how equilibrium opinion deviate from the average, while disagreement measures the total difference of equilibrium opinions between each pair of individuals.
There is an tradeoff between polarization and disagreement, and the polarization-disagreement index (PDI) is defined, which is the sum of polarization and disagreement.
We also review works of controlling polarization and disagreement by manipulating individuals' internal opinions or network structure.
These problems can be efficiently solved by convex optimization or greedy algorithm.

Although the polarization and disagreement problems have been studied, there are still some issues need to be investigated.
First, most of the control strategies are based on the polarization and disagreement equilibrium opinion.
In reality, it is often expected to control polarization and disagreement as soon as possible.
Therefore, it is interesting to study how to control polarization and disagreement dynamically.
Second, the evaluation and control are based on the Fredkin-Johnsen model.
However, ingredients like noise effect, external influence is common in reality.
The influence of these ingredients on opinion polarization and disagreement, and how to control polarization and disagreement in such settings are also need to be exploited.
Last but not least, there might exist attackers who want to maximize the polarization and disagreement, while some defenders expected to minimize them.
It is also interesting to investigate the evolution of polarization and disagreement in such adversarial setting.

\section{Acknowledgements}
This research is supported by This work is supported by the National Key Research and Development Program of China (2017YFB1400100).

\bibliography{LYJ_Trans}
\end{document}